\documentclass[preprint,showpacs,12pt,preprintnumbers,amsmath,amssymb,nofootinbib]{revtex4}
\pdfoutput=1
\usepackage{bbm}
\usepackage{amsfonts}
\usepackage{booktabs}
\usepackage{mathrsfs}
\usepackage{epsfig}
\usepackage{graphicx}
\usepackage{dcolumn}
\usepackage{bm}
\usepackage{amsmath}
\usepackage{color}
\usepackage{slashed}
\usepackage{subfigure}

\let\jnfont=\rm
\def\NPB#1,{{\jnfont Nucl.\ Phys.\ B }{\bf #1},}
\def\PLB#1,{{\jnfont Phys.\ Lett.\ B }{\bf #1},}
\def\EPJC#1,{{\jnfont Eur.\ Phys.\ Jour.\ C }{\bf #1},}
\def\PRD#1,{{\jnfont Phys.\ Rev.\ D }{\bf #1},}
\def\PRL#1,{{\jnfont Phys.\ Rev.\ Lett.\ }{\bf #1},}
\def\MPLA#1,{{\jnfont Mod.\ Phys.\ Lett.\ A }{\bf #1},}
\def\JPG#1,{{\jnfont J.\ Phys.\ G}{\bf #1},}
\def\CTP#1,{{\jnfont Commun.\ Theor.\ Phys.\ }{\bf #1},}
\def\ZPC#1,{{\jnfont Z.\ Phys.\ C }{\bf #1},}
\def\JHEP#1,{{\jnfont JHEP \ }{\bf #1},}
\def\Rv{\not{\hbox{\kern-1pt $R$}}}
\def\p{\not{\hbox{\kern-3pt $p$}}}

\begin{document}

\title{Non-minimal Higgs-gravity coupling and naturalness at the LHC and beyond}

\author{ Lei Wu$^{1}$}
\author{ Huaying Zhang$^{2}$}
\author{ Bin Zhu$^3$}

\affiliation{
 $^1$ Department of Physics and Institute of Theoretical Physics, Nanjing Normal University, Nanjing, 210023, China\\
 $^2$ College of Physics and Material Science, Henan Normal University, Xinxiang, 453007, China\\
 $^3$ Department of Physics, Yantai University, Yantai 264005, China
 }%

\begin{abstract}
The non-minimal Higgs-gravity coupling, $\xi H^\dagger H R$, can play an interesting role in particle physics. We utilize the available LHC Run-2 data to obtain the bound on the coupling $\xi$ and discuss the prospects of constraining it at CEPC, FCC-ee and ILC. Also, we investigate its implication on the naturalness.
\end{abstract}

\maketitle

\section{Introduction}
The discovery of the Higgs boson at the LHC completes the Standard Model (SM). It is a great step towards understanding the electroweak symmetry breaking mechanism. Up to now, most of measurements are compatible with the SM predictions at $1-2 \sigma$ level. Search for new physics beyond the SM will be next essential task in the LHC experiments. However, there are still no convincing deviations from the SM reported in on-going particle physics experiments. This drives us to reconsider the question of the scale of new physics.

As well known, Planck energy $M_{pl}=\sqrt{8\pi G} \simeq 10^{18}$ GeV is a definitely higher physics scale, where the gravity that is not included in the SM will become important. By far, general relativity (GR) is the best description of gravity in terms of geometric language. However, GR cannot be the correct theory at the quantum level as it is not renormalizable. From the prospective of effective field theory, GR can be considered as the leading order term of a more fundamental
high energy theory that is diffeomorphism invariant, which is given by
\begin{align}
S_{\text{EH}}=\int d^4 x\sqrt{-g} \left[\frac{M_{pl}^2}{2}(R-2\Lambda)+c_1 R^2+c_2 R_{\mu\nu}R^{\mu\nu}+{\cal O}(M_{pl}^{-2})\right].
\label{eh}
\end{align}
Here $R$ and $R_{\mu\nu}$ are Recci curvature scalar and tensor. $\Lambda$ denotes the cosmological constant. $c_1$ and $c_2$ are dimensionless parameters. Higher dimensional operators are neglected in Eq.~(\ref{eh}), because they are suppressed by increasing powers of the Planck scale $M_{pl}$.

Assuming the SM is a valid theory between electroweak and planck scale, the SM particles will interact with the gravity. The matter fields in GR couple to gravity through their energy-momentum tensor in the manner of minimal coupling,
\begin{align}
S_{\text{MC}}=-\int d^4 x\sqrt{-g}\delta g_{\mu\nu} T^{\mu\nu}_{\rm SM},
\end{align}
where $T^{\mu\nu}_{\rm SM}$ is the energy-momentum tensor. Besides, up to dimension-4 interactions, the Higgs doublet $H$ can also couple to the curvature scalar $R$ via an unique dimension-4 operator,
\begin{align}
S_{\text{NMC}}=-\int d^4 x\sqrt{-g} \xi R H^{\dagger}H,
\end{align}
where $\xi$ is a dimensionless parameter. Such a non-minimal coupling was first pointed out in reference~\cite{Chernikov:1968zm} and then in the study of quantization of a scalar field on a curved spacetime~\cite{Callan:1970ze}. It can play an important role in Higgs inflation, see e.g.~\cite{Bezrukov:2007ep}. On the other hand, this non-minimal coupling could affect the stabilization of the electroweak scale. The quadratic divergence of the Higgs mass can be counterbalanced with the quartic divergence of the vacuum energy that depends on $\xi$. Besides, it will lead to the change of Higgs couplings with the SM particles and can be constrained by the LHC data~\cite{Atkins:2012yn} and unitarity bound~\cite{Ren:2014sya,Xianyu:2013rya}.

In this work, we will first update the bound on the coupling $\xi$ with the available LHC Run-2 data in Sec.~\ref{bound} and then discuss implications of this non-minimal coupling on naturalness in Sec.~\ref{naturalnss}. Finally, we draw our conclusions in Sec.~\ref{conclusion}.

\section{Bounds on no-minimal coupling at the LHC and beyond}\label{bound}

In the Jordan frame, the action including non-minimal coupling can be written as,
\begin{align}
S_{J}&=\int d^4 x\sqrt{-g^J}\left(\frac{1}{2}M^2+\xi H^{\dagger}H\right)R^J-\frac{1}{4}\sum_j F_{\mu\nu j}F_{j}^{\mu\nu}\nonumber\\
        &+(D_{\mu}H)^{\dagger}(D^{\mu}H)-V(H),
        \label{action}
\end{align}
where $H$ is the Higgs doublet and $F$ is the gauge field strength tensor.  For simplicity, we ignore the cosmological constant $\Lambda$ in Eq.~(\ref{action}). The corresponding equation of motion after differentiating with action is given by,

\begin{align}
\left\{R_{\mu\nu}^J-\frac{1}{2}g_{\mu\nu}^J R^J+(g_{\mu\nu}^J\Delta_{\lambda}\Delta^{\lambda}-\Delta_{\mu}\Delta_{\nu})\right\}\Omega^2=-\frac{1}{M_P^2}T_{\mu\nu},
\end{align}
where the dimensionless parameter $\Omega$ is defined as,
\begin{align}
\Omega^2=\frac{M^2+2\xi H^{\dagger}H}{M_{P}^2}.
\end{align}
In order to recover canonical Einstein equation, the canonical Higgs kinetic term must be renormalized, which can be realized by the Weyl transformation,
\begin{align}
g_{\mu\nu}^J &=\Omega^{-2}g_{\mu\nu}\nonumber\\
g^{\mu\nu J} &=\Omega^{2}g^{\mu\nu}\nonumber\\
\sqrt{-g^J}&=\Omega^{-4}\sqrt{-g}\nonumber\\
R^{J}&=\Omega^2(R-6\Delta^{\lambda}\Delta_{\lambda}\log\Omega+
6g^{\mu\nu}\Delta_{\mu}\log\Omega\Delta_{\nu}\log\Omega)
\end{align}

Then the action in the Einstein frame can be rewritten as follows,
\begin{align}
S_E&=\int d^4 x\sqrt{-g}(-\frac{1}{2}M_P^2 R-\frac{1}{4}\sum_j F_{\mu\nu j}F_{j}^{\mu\nu}\nonumber\\
       &+\frac{1}{\Omega^2}(D_{\mu}H)^{\dagger}(D^{\mu}H)-\frac{1}{\Omega^4}V(H)
       +\frac{3\xi^2}{M_P^2\Omega^4}(\nabla_{\mu}H^{\dagger}H)^2)
\end{align}
Besides, the additional transformation of the Higgs doublet is required to obtain the canonical Einstein equation,
\begin{align}
\chi=\zeta H=\frac{1}{\sqrt{1+\beta}}H
\label{higgs}
\end{align}
where $\beta=6\xi^2 v^2 / M_P^2$. We can see that the wave function of new Higgs field $\chi$ is suppressed by a factor of $\beta$. This will lead to the change of the Higgs couplings and affect the observed Higgs signal strength at the LHC. We should mention that the wavefunction renormalization effect in eqn~\ref{higgs} is not only in Einstein frame but in Jordan frame. The reason why we focus on Einstein frame is that it is convenient to deal with Higgs detection.

\begin{table}
\caption{Expected statistical precisions, $\Delta(\sigma\cdot Br)/(\sigma\cdot Br)$, for Higgs measurements at the HL-LHC(14 TeV, 3 ab$^{-1}$)~\cite{Dawson:2013bba}, ILC(250 GeV, 2 ab$^{-1}$)~\cite{Barklow:2015tja}, FCC-ee(240 GeV, 5 ab$^{-1}$)~\cite{Gomez-Ceballos:2013zzn} and CEPC(240 GeV, 5 ab$^{-1}$)~\cite{cepc}.}
\begin{center} \begin{tabular}{|p{2.5cm}<{\centering}|p{2.2cm}<{\centering}|p{2.2cm}<{\centering}|p{2.2cm}<{\centering}|p{2.5cm}<{\centering}|}\hline
			Facility          &HL-LHC   &ILC    &CEPC        &FCC-ee  \\
			\hline
			$\kappa_{\gamma}$ &$2-5\%$  &12\%   &9.0\%       &4.2\% \\
			\hline
			$\kappa_{g}$      &$3-5\%$  &2.5\%  &1.6\%       &1.98\%  \\
			\hline
			$\kappa_{W}$      &$2-5\%$  &1.1\%  &1.5\%       &1.27\%   \\
			\hline
			$\kappa_{Z}$      &$2-4\%$  &6.7\%   &4.3\%      &4.4\%   \\
			\hline
			$\kappa_{u}$      &$7-10\%$ &$-$    &$-$         &$-$   \\
			\hline
			$\kappa_{d}$      &$4-7\%$  &$-$    &$-$         &$-$    \\
			\hline
			$\kappa_{c}$      &$7-10\%$ &2.9\%  &2.2\%       &1.7\%     \\
			\hline
			$\kappa_{s}$      &$4-7\%$  &$-$    &$-$         &$-$  \\
	     	\hline
	     	$\kappa_{t}$      &$7-10\%$ &$-$    &$-$         &$-$     \\
	     	\hline
		    $\kappa_{b}$      &$4-7\%$  &0.42\%  &0.28\%     &0.28\% \\
		    \hline
    	\end{tabular}	
	\end{center}
\label{tab:precision}
\end{table}
Up to now, various productions and decays of Higgs boson have been measured at the LHC.  The signal strength of a single Higgs boson in a specific analysis can be given by,
\begin{equation}
\mu = \sum_{i} c_i\omega_i,
\label{Eq:mu}
\end{equation}
where the sum runs over all production and decay channels used in the analysis. For $i$-th channel, the signal strength can be calculated by
\begin{equation}
c_i=\frac{\left[\sigma\times BR\right]_i}{\left[\sigma_{SM}\times BR_{SM}\right]_i},
\label{Eq:ci}
\end{equation}
and the corresponding weight is given by,
\begin{equation}
\omega_i=\frac{\epsilon_i\left[\sigma_{SM}\times BR_{SM}\right]_i}{\sum_j\epsilon_j\left[\sigma_{SM}\times BR_{SM}\right]_j}.
\label{Eq:omega}
\end{equation}
Here $\epsilon_i$ is the relative experimental efficiencies for $i$-th channel. We confront the modified Higgs couplings with the available LHC Higgs data and calculate the $\chi^2$ with the package ~\textsf{HiggsSignals-2.2.0}~\cite{higgssignals}, which includes 102 observables from the LHC run-1 and 2. The number of degrees of freedom $n_{dof}$ is 101. We take the Higgs mass $m_h=125$ GeV and choose the peak-centered $\chi^2$ method to perform the fit.

\begin{figure}[ht]
	\centering
	\includegraphics[width=3.5in,height=3in]{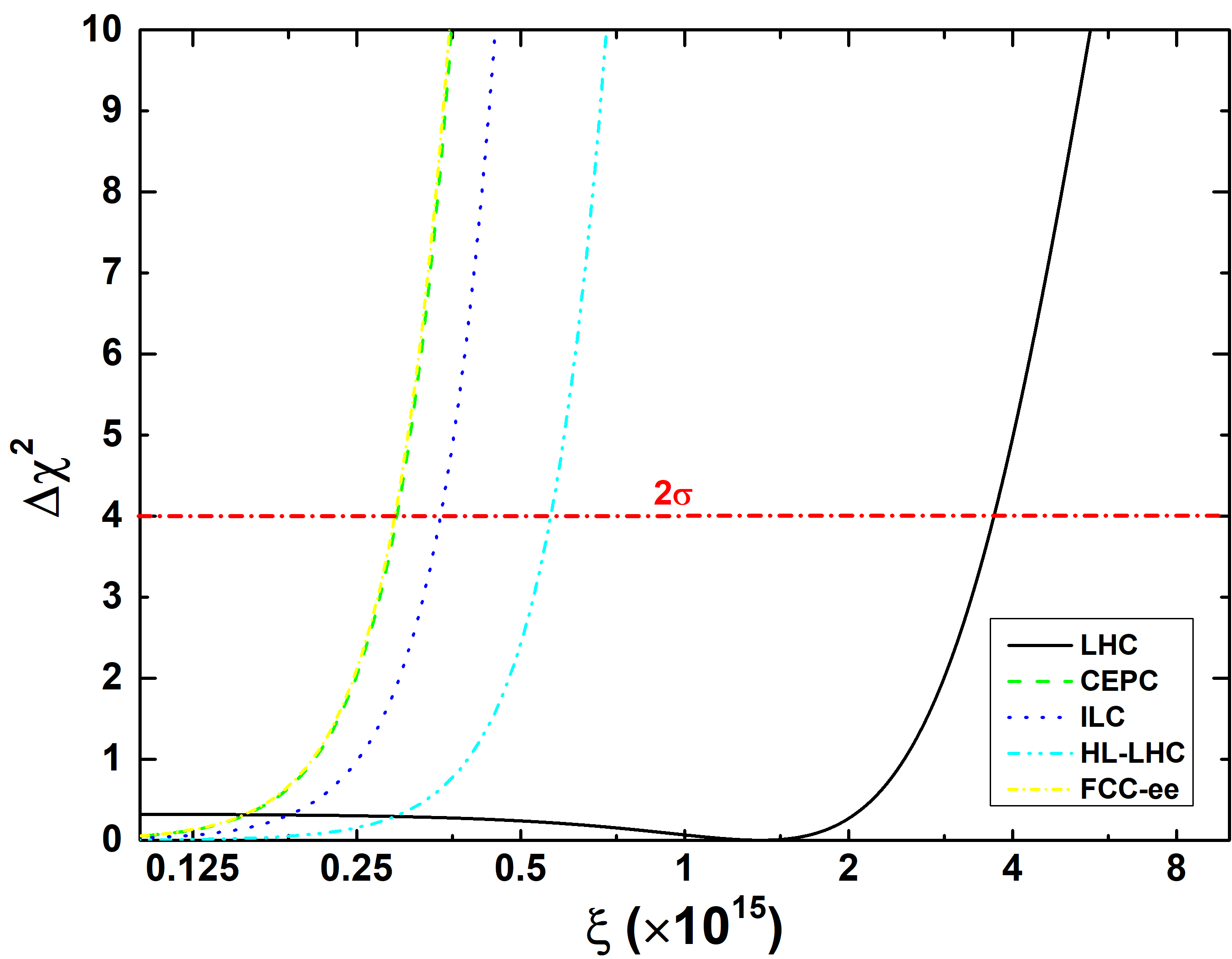}\vspace{-0.5cm}
	\caption{$\Delta\chi^2$ distribution for the non-minimal coupling $\xi$. The black solid line corresponds to the fitting result with current LHC data. The green dashed (CEPC), blue dotted (ILC), cyan dash-dot-dot (HL-LHC) and yellow dash-dot lines (FCC) corresponds the fitting results with the expected precisions in Table~\ref{tab:precision}, respectively. The red dash-dot line denotes the $2\sigma$ upper bound of $\xi$.}
	\label{tlep}
\end{figure}
In Fig.~\ref{tlep}, we show the $\Delta\chi^2$ distribution in one dimension for the non-minimal coupling $\xi$. For plotting distributions of $\Delta\chi^2$ as a function of one variable, the $2\sigma$ exclusion limit is then given by $\Delta\chi^2 = 4$. The appearance of additional factor $\zeta$ in Eq.~(\ref{higgs}) will result in an overall suppression of Higgs coupling with the SM particles, which will be constrained by the Higgs measurements. It can be seen that $\xi > 3.7 \times 10^{15}$ has been excluded by the current Higgs data at $2\sigma$ level. Besides, we also estimate the prospects of constraining the non-minimal coupling $\xi$ in future HL-LHC, ILC, FCC-ee and CEPC experiments. In the fitting, we assume that all the measured Higgs couplings will be the same as the SM couplings with the expected measurement uncertainties ($\sigma_i$) given in Tab.~\ref{tab:precision}. By calculating $\chi^2=\displaystyle{\sum_{i}^{N}}(\mu_i-1)^2/\sigma^2_i$, we find that the HL-LHC will be able to improve the current LHC upper limit of $\xi$ by a factor of 5. More importantly, the future lepton colliders, such as CEPC and FCC-ee, will further reduce this upper bound by an order.

It should be noted that the cut-off $\Lambda$ in our scenario is proportional with $1/\xi$ given by the tree-level unitarity bound~\cite{Burgess:2009ea}. The allowed large ${\cal O}(\xi) \sim 10^{15}$ may lead to a low cut-off, i.e. ${\cal O}(\Lambda) \sim 10^3$ GeV, which seems inconsistent with the current LHC data and may have the unitarity problem. However, in~\cite{Aydemir:2012nz,Calmet:2013hia}, the authors studied the perturbative unitarity violation in the standard Higgs inflation model and found that the Cutkosky cutting rule implied by perturbative unitarity is fulfilled at one-loop. This strongly indicated that unitarity can be restored order by order in perturbation theory so that there is no need to worry about the low cut-off problem. Such a mechanism is known as self-healing mechanism. Therefore, the Standard Model with a non-minimal coupling of the Higgs boson to space-time curvature can describe particle physics and inflation up to Plank scale. On the other hand, the large coupling $\xi$ may re-introduces the hierarchy problem for Higgs boson, since its quadratic divergence is proportional to large cut-off $\Lambda^2$. But in the next section, we will show the large $\xi$ plays a central role in reducing the fine-tuning.



\section{Implication on Naturalness}\label{naturalnss}
In the SM, the Higgs mass $m_h$ is proportional to vacuum expectation value $v$, $m^2_h=\lambda_h v^2$, where $\lambda_h$ is Higgs trilinear self-coulping. The appearance of non-minimal coupling will lead to the vacuum energy $V_0$ contributing to this relation and affect the quantum corrections to the Higgs mass~\cite{Demir:2014gca}. To see this, we start from the Einstein equation in a spontaneously broken phase background, which is given by
\begin{align}
&R=\frac{1}{M_{P}^2-\xi v^2}(4V_0+2m_h^2 v^2+\lambda_h v^4),\\
&m_h^2+\lambda_h v^2+\xi R=0. \label{mh}
\end{align}
When $\xi=0$, Eq.~(\ref{mh}) will recover the SM tree-level relation of vacuum expectation value and Higgs mass. Then, we can obtain the new tree-level vev straightforwardly,
\begin{align}
 v^2=\frac{-m_h^2 M_P^2-4V_0\xi}{M_P^2\lambda_h+m_h^2\xi}
 \label{eqn:tree}
\end{align}
It should be noted that the quadratic and quartic divergences will appear when we calculate the one-loop correction to the cosmological constant and Higgs mass~\cite{Deser:1974zzd,Deser:1974nb,Demir:2014gca},
 \begin{align}
 \delta V_0&=\frac{1}{16\pi^2}\left(\frac{1}{4}(n_F-n_B)\Lambda^4+2m_h^2\Lambda^2\right)\nonumber\\
 \delta m_h^2&=\frac{3}{16\pi^2}(2\lambda_h+\frac{g_Y^2}{4}+\frac{3g_2^2}{4}-2y_t^2)\Lambda^2
 \label{eqn:loop}
  \end{align}
After expanding terms up to ${\cal O} (\Lambda^2/M_P^2)$, we have the one-loop correction to vev as followings,
 \begin{align}
 \delta v^2\sim \left(1-\frac{\Lambda^2}{M_P^2}\xi\right)\Lambda^2
 \label{shiftvev}
 \end{align}
It can be seen that $\delta v^2$ may vanish for a specific value of $\xi$ so that the fine-tuning problem can be simply removed.
In order to quantitatively calculate the fine-tuning, we define the following measure $\Delta$,
\begin{equation}
\Delta=\delta v^2/v^2.
\end{equation}


\begin{figure} [ht]
\begin{center}
\includegraphics[width=4.0in,height=3.5in]{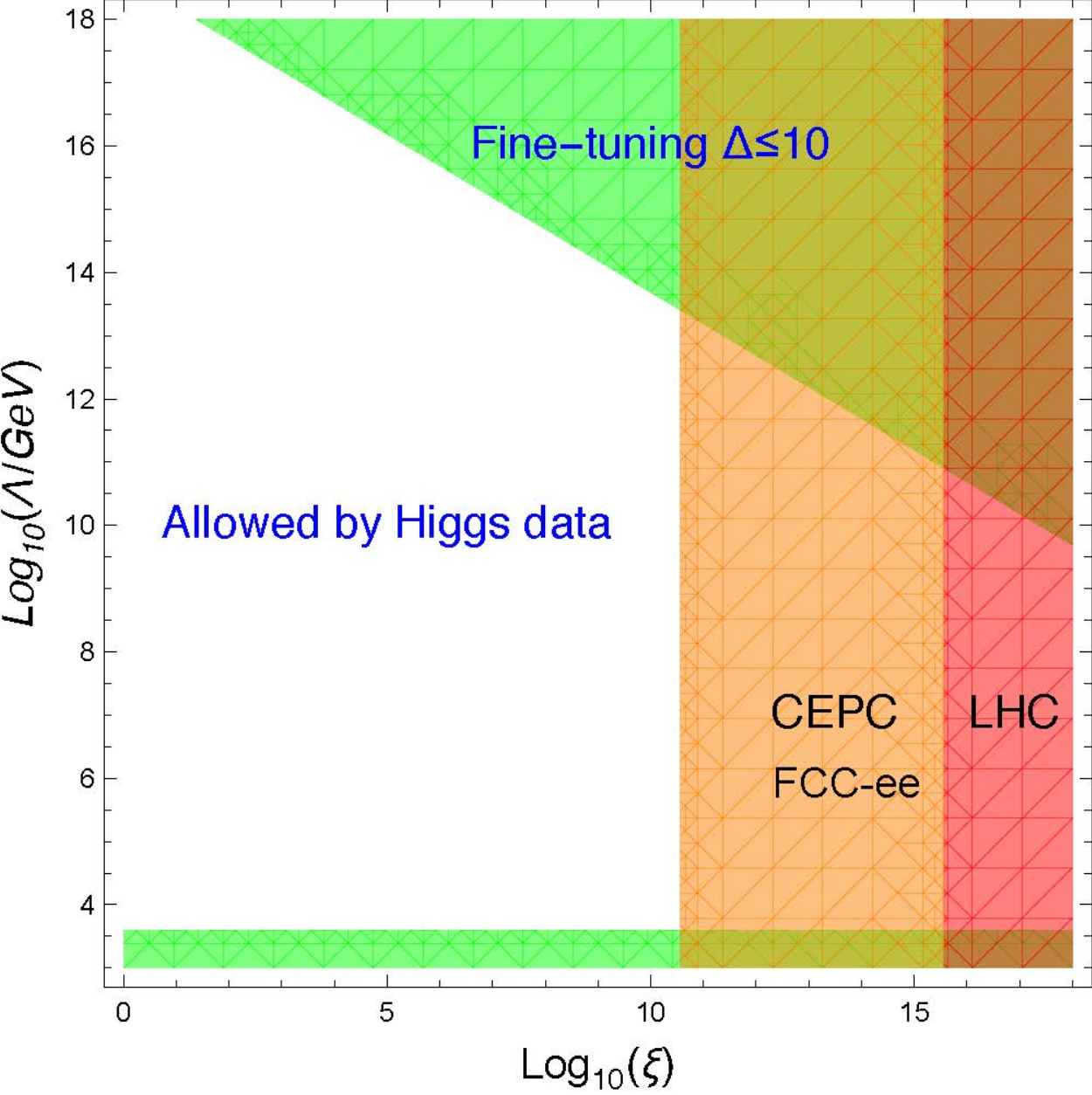}
\end{center}
\caption{Region plot of fine-tuning measure $\Delta \leq 10$ (green color) on the plane of $\xi$ versus UV cut-off $\Lambda$. Red and orange regions have been excluded by the current LHC Higgs data and future Higgs measurements at CEPC/FCC-ee, respectively.}
\label{finetot}
\end{figure}
In Fig.~\ref{finetot}, we show region plot of fine-tuning measure $\Delta \leq 10$ (green color) on the plane of $\xi$ versus UV cut-off $\Lambda$. We can see that there are two separated green regions. The lower region is trivial because the UV cut-off $\Lambda$ is close to TeV scale. While the upper region is due to the cancelation effect of $\xi$ in Eq.~(\ref{shiftvev}). The red and orange regions are from the bounds given by current LHC Higgs data and expected Higgs measurements at CEPC/FCC-ee, which can already exclude a part of region with $\Delta \leq 10$. It should be noted that the UV cut-off $\Lambda$ is  arbitrarily large since the unitarity contion $\Lambda \lesssim M_P/\xi$~\cite{Burgess:2009ea} is restored automatically.  In other words, the higher dimensional operator is highly suppressed by the large cut-off which plays no role in Higgs boson quantum corrections. We should mention that our stabilization of hierarchy between Plank scale and weak scale is achieved at one-loop level. This situation also happens in strong dynamics approaches to solve hierarchy problem, such as Composite Higgs and Twin Higgs models. The quadratic divergence in our framework will reappear in higher order corrections, which may be cancelled by adjusting the value of $\xi$ technically. 
\section{Conclusion}\label{conclusion}
In this paper, we confronted the non-minimal Higgs-gravity coupling $\xi H^\dagger H R$ with the LHC Higgs data and investigate its implication on naturalness as the cancellation between quadratic divergence of the Higgs mass and the quartic divergence of the vacuum energy. We found that the dimensionless parameter $\xi > 3.7 \times 10^{15}$ has been excluded at 95\% C.L.., which constrained the low fine-tuning region with large coupling $\xi$. The upper bound on $\xi$ will be further reduced to $2.9 \times 10^{-14}$ at CEPC and FCC-ee.

\acknowledgments
This work is supported by the National Natural Science Foundation of China (NNSFC) under grants No. 11805161, 11747026 (BZ) and 11705093 (LW). BZ is also supported by Natural Science Foundation of Shandong Province under grant No. ZR2018QA007. LW is also supported by Jiangsu Specially Appointed Professor Program.

\end{document}